\documentclass[nocompress]{spie}  %>>> use for US letter paper
\usepackage{amsmath,amsfonts,amssymb}
\usepackage{svg}

\usepackage{graphicx}
\usepackage{txfonts}
\usepackage[colorlinks=true,citecolor=blue]{hyperref}
\usepackage{xspace}
\usepackage{array}
\usepackage{bm}% to write bold letters in math mode
\usepackage{color}
\usepackage{mathtools}
\usepackage{enumerate}
\usepackage{color}
\usepackage{ulem}
\usepackage{appendix}
\usepackage[stable]{footmisc}
\usepackage{relsize}
\usepackage{aas_macros}

\newcommand\arcsec{\mbox{$^{\prime\prime}$}}%
\newcommand\farcs{\mbox{$.\!\!^{\prime\prime}$}}%

\title{Design, scientific goals, and performance of the SCExAO survey for planets around accelerating stars}

\author[a]{Mona El Morsy}
\author[a,b]{Thayne Currie}

\author[c,d]{Masayuki Kuzuhara}
\author[e]{Jeffrey Chilcote}

\author[b,c,f,g]{Olivier Guyon}
\author[h]{Taylor L. Tobin} %ORCID 0000-0001-8103-5499, if needed
\author[i]{Timothy Brandt}
\author[j]{Qier An}
\author[b]{Kyohoon Anh}
\author[a]{Danielle Bovie}
\author[b]{Vincent Deo}
\author[t]{Tyler Groff}
\author[k]{Ziying Gu}
\author[l]{Markus Janson}
\author[u]{Nemanja Jovanovic}
\author[h]{Yiting Li}
\author[t]{Kellen Lawson}
\author[b]{Julien Lozi}
\author[y]{Miles Lucas}
\author[m]{Christian Marois}
\author[n]{Naoshi Murakami}
\author[o]{Eric L. Nielsen}
\author[v,w]{Barnaby Norris}
\author[p]{Nour Skaf}
\author[k, c, d]{Motohide Tamura}
\author[m]{William Thompson}
\author[r,c]{Taichi Uyama}
\author[b]{Sebastien Vievard}

\affil[a]{Department of Physics and Astronomy, University of Texas-San Antonio, San Antonio, TX 78006 USA}
\affil[b]{National Astronomical Observatory of Japan, Subaru Telescope, 650 North A'oh\=ok\=u Place, Hilo, HI 96720, U.S.A.}
\affil[c]{Astrobiology Center of NINS, 2-21-1 Osawa, Mitaka, Tokyo 181-8588, Japan}
\affil[d]{National Astronomical Observatory of Japan, Mitaka, Tokyo 181-8588, Japan}
\affil[e]{Department of Physics and Astronomy, University of Notre Dame, 225 Nieuwland Science Hall, Notre Dame, IN 46556 USA}
\affil[f]{Steward Observatory, University of Arizona, Tucson, AZ 85721, USA}
\affil[g]{Wyant College of Optical Sciences, University of Arizona, Tucson, AZ 85721, USA}
\affil[h]{Department of Astronomy, University of Michigan, 1085 S. University, Ann Arbor, MI 48109, USA}
\affil[i]{Space Telescope Science Institute, Baltimore, MD 21218, USA}
\affil[j]{Department of Physics, University of California, Santa Barbara, Santa Barbara, CA 93106, USA}
\affil[k]{Department of Astronomy, Graduate School of Science, The University of Tokyo, 7-3-1, Bunkyo-ku, Tokyo 113-0033, Japan}
\affil[l]{Department of Astronomy, Stockholm University, Stockholm 10691, Sweden}
\affil[m]{National Research Council Herzberg Astronomy and Astrophysics, Victoria, B.C. V9E 2E7, Canada}
\affil[n]{Faculty of Engineering, Hokkaido University, Kita 13, Nishi 8, Kita-ku, Sapporo, Hokkaido 060-8628, Japan}
\affil[o]{Department of Astronomy, New Mexico State University, 1320 Frenger Mall, Las Cruces, NM 88003, USA}
\affil[p]{Department of Astronomy \& Astrophysics, University of California, Santa Cruz, CA 95064, USA}
\affil[r]{Department of Physics and Astronomy, California State University Northridge, 18111 Nordhoff Street, Northridge, CA 91330, USA}
\affil[t]{NASA-Goddard Space Flight Center, Greenbelt, MD, USA}
\affil[u]{Department of Astronomy, California Institute of Technology, 1200 E. California Blvd.,Pasadena, CA, 91125, USA}
\affil[v]{Sydney Institute for Astronomy, School of Physics, Physics Road, University of Sydney, NSW 2006, Australia Sydney}
\affil[w]{Astrophotonic Instrumentation Laboratories, Physics Road, University of Sydney, NSW 2006, Australia}
\affil[y]{Institute for Astronomy, University of Hawaii-Manoa, Honolulu, HI, USA}
\authorinfo{Send correspondence to M. El Morsy (\href{mailto:mona.elmorsy@utsa.edu}{mona.elmorsy@utsa.edu}).}

\begin{document} 
\maketitle

\begin{abstract}
We describe the motivation, design, and early results for our 42-night, 125 star Subaru/SCExAO direct imaging survey for planets around accelerating stars.   Unlike prior large surveys, ours focuses only on stars showing evidence for an astrometric acceleration plausibly due to the dynamical pull of an unseen planet or brown dwarf.  Our program is motivated by results from a recent pilot program that found the first planet jointly discovered from direct imaging and astrometry and resulted in a planet and brown dwarf discovery rate substantially higher than previous unbiased surveys like GPIES.  The first preliminary results from our program reveal multiple new companions; discovered planets and brown dwarfs can be further characterized with follow-up data, including higher-resolution spectra.  Finally, we describe the critical role this program plays in supporting the Roman Space Telescope Coronagraphic Instrument, providing a currently-missing list of targets suitable for the CGI technological demonstration without which the CGI tech demo risks failure.
%and SCExAO/CHARIS spectra and complementary Keck/NIRC2 thermal infrared photometry constshowing dynamical evidence for a companionThis 42-night, 125 star survey uses the leading planet imaging system in the northern hemisphere -- SCExAO/CHARIS -- and complementary Keck/NIRC2 data to discover and characterize the atmospheres of planets around stars show
%\textcolor{red}{TC - rewrite this}
%Ground-based ExAO surveys (GPIES and SHINE) have detected about 20 planets by direct imaging. However, the overall discovery rate of these `blind’ surveys is low. Direct imaging data often poorly constrain planet orbits and cannot directly measure planet masses and thus are limited in revealing how planet atmospheres evolve as a function of mass. We describe the direct imaging survey: a search for planets around accelerating stars using Subaru Coronagraphic Extreme Adaptive Optics (SCExAO) project coupled with the CHARIS integral field spectrograph, discuss post-processing approaches, tradeoffs in planet atmospheric characterization between low (R $\sim$20) and higher (R $\sim$70) resolution data, and detection sensitivities. We detail our understanding of the atmospheric evolution of gas giant planets and forecast the discovery yield and scientific insights gained from a larger, 40-night Intensive Survey with an upgraded SCExAO. We describe how this survey retires a key technical risk with the Roman CGI technological demonstration by providing suitable targets: a critical component that is currently missing.

\end{abstract}


%\newpage
\section{Introduction}
\label{sec:intro}  
Direct imaging is an observing technique that is well suited for detecting young Jupiter-like planets and brown dwarfs \cite{CurrieBiller2023}, provides key information about exoplanetary atmospheres \cite{Currie2011,Barman2015}, and characterizing the atmospheres and will eventually confirm and characterize Earth-like planets around nearby stars.  Only about 20-25 extrasolar planets have been discovered this way, a small sample compared to thousands found via radial velocities and transits \cite{CurrieBiller2023}.   Detected companions are typically more than 5 M$_{J}$ in mass and orbit beyond 20–30 au, far from the peak jovian planet frequency around 3 au \cite{Fulton2021}.  

Most direct imaging discoveries thus far draw from so-called blind (i.e. unbiased) surveys, where targets are selected based on system properties like age and distance. However, the low yields of these blind surveys have shown exoplanets detectable using current direct imaging instruments are rare (e.g. \cite{Nielsen2019,Vigan2021}). Absent enormous, unfeasible contrast gains from the ground in the next few years enabling reflected-light planet detection -- i.e. 10$^{-8}$ at $<$0.5" in the near IR -- any \textit{blind} survey conducted prior to the advent of extreme AO systems on 20-30m class extremely large telescopes (ELTs) will also have low yields.

Furthermore, direct imaging by itself has some key limitations for exoplanet characterization.  While direct imaging provides exceptional constraints on an exoplanet’s atmospheric properties, it does not directly measure a planet's mass. Masses typically reported for directly-imaged planets are \textit{inferred} through evolutionary model predictions that map between a planet's luminosity and mass as a function of age (e.g. \cite{Baraffe2003}).  But this mapping is highly uncertain, especially at young ages due where planets are brightest compared to their host stars \cite{SpiegelBurrows2012}.   
Additionally, the typically wide separations and short temporal coverage for the locations of imaged exoplanets around their host stars can also lead to poor constraints on orbital parameters derived purely from direct imaging data alone \cite{Bowler2020}.

%The low number of direct imaging detections is due to the small percentage of imaged exoplanets in unbiased surveys. Even with advanced extreme adaptive optics on 8-10m class telescopes, detection frequencies are only a few percent around AFGK stars \cite{Nielsen2019}.

%Ground based telescopes with adaptive optics (AO) have now provided direct imaging detections of about 20 extrasolar gas giant planets, largely drawing from so-called blind (i.e. unbiased) surveys, where targets are selected based on system properties like age and distance. However, the low yields of these blind surveys have shown exoplanets detectable using current direct imaging instruments are rare.  While direct imaging provides exceptional constraints on an exoplanet’s atmospheric properties, it does not directly measure a planet's mass and can often give only weak constraints on a planet's orbit. The dearth of direct imaging detections and poor mass and orbital constraints for imaged companions impede our understanding of gas giant (i.e. jovian) exoplanet atmospheres, planet evolution, and the architectures of planetary systems.

Instead of a blind search, direct imaging campaigns coupled with another indirect detection method sensitive to the planet's dynamical influence can improve discovery yields, provide better constraints on the planet's orbit, and directly estimate the planet's mass.  In particular, monitoring of a star's \textbf{astrometry} -- i.e. its proper motion across the sky -- can identify which those that are undergoing a proper motion
acceleration caused by an unseen planet\footnote{Radial-velocity (RV) data can also identify stars being gravitationally perturbed by a companion.  However, RVs are typically ill suited for identifying planets around the stars with the highest frequency of imaged planets (main sequence B, A, and early F stars); RV data is generally less precise for young, more chromospherically-active Sun-like stars than for Gyr-old solar twins. }.
By jointly analyzing an imaged planet’s relative astrometry from imaging and the host star’s absolute astrometry can yield precise, directly-determined
planet masses and improved constraints on orbital properties \cite{GMBrandt2021}. The micro-arcsecond precision of the European Space Agency’s \textit{Gaia} mission combined with measurements 25 years
prior from Hipparcos is sufficient to enable the astrometric detection of superjovian planets at
Jupiter-to-Neptune like separations around the nearest stars.

Our Subaru and Keck direct imaging survey\footnote{We refer to this program in general terms here.  The program's official name and acronym will be announced later.} takes a different approach to discovering and characterizing exoplanets:
\begin{enumerate}
    \item We select 175 young, nearby stars for direct imaging observations based on dynamical evidence for a companion from precision astrometry contained in the Hipparcos-Gaia Catalogue of Accelerations (HGCA): i.e. stars showing an astrometric acceleration.
    \item We target these accelerating stars using the leading planet-imaging system in the northern hemisphere -- the Subaru Coronagraphic Extreme Adaptive Optics Project (SCExAO) coupled with the CHARIS integral field spectrograph \cite{Jovanovic2015,Groff2016} -- and (for the brightest planets and brown dwarfs) obtain complementary thermal infrared (IR) imaging with NIRC2 camera on the Keck II Telescope.
   % \item Direct imaging data gives us information about a planet's atmosphere; imaging + astrometry together can constrain the planet's dynamical mass and orbit.  Thus, 
    \item We use the \texttt{orvara} dynamical code to simultaneously constrain the planets' masses and orbits \cite{Brandt2021}.  Empirical libraries and new, sophisticated atmospheric models will constrain the planets' atmospheres.
\end{enumerate}

The likely result of this program will be a planet/low-mass brown dwarf discovery rate 5 times higher than that of blind imaging surveys and new benchmark sample of exoplanets that are imaged, weighed, and have their orbits tracked.  Our discoveries will anchor models of substellar formation and evolution from the largest brown dwarfs to jovian exoplanets.   This sample will have critical long-term value, by providing targets and thus retiring risk for the \textit{Nancy Grace Roman Space Telescope} technology demonstration experiment.

\section{Survey Allocation and Size}
\label{sec:SO}

This program consists of guaranteed telescope time: 42 nights split between SCExAO/ CHARIS (34 nights) and Keck/NIRC2 (8 nights) between February 2024 and July 2026.
The telescope time derives from two separate allocations. The first allocation draws from an Intensive Survey proposal through the Subaru time allocation committee consisting of 32 SCExAO/CHARIS nights between February 2024 and July 2026 (Program S24-023I; PI. T. Currie).   The second is a NASA Keck Strategic Mission Support (KSMS) proposal for eight nights of Keck/NIRC2 time and two nights of SCExAO/CHARIS time through the Keck/Subaru time exchange for nights between February 2024 and January 2026 (Program 2024A$\_$N004, PI. T. Currie).

For reasons described below, our parent sample consists of about 175 stars and we expect to observe about 125 of them, considering expected weather losses and time needed to confirm candidate planets and brown dwarfs.   Our sample is intermediate in size between GPIES\cite{Nielsen2019} and other recent surveys such as the LBT/LEECH campaign \cite{Stone2018}. 
%somewhat less than half the size of GPIES 
%is about 1/4th that of GPIES, The total sample size is about a we plan to nominally observe up to 5 targets per night.  Considering weather losses and time needed to %Our survey derives from two separate time allocations

%Our Intensive Survey targets nearby young accelerating stars, primarily using the Subaru Coronagraphic Extreme Adaptive Optics project (SCExAO) coupled with the CHARIS integral field spectrograph, which covers wavelength from 1.1—2.4 $\mu m$ with Keck/NIRC2 for follow-up in the thermal infrared at 3.8 $\mu m$.  The survey is composed of 42 nights from Semester 2024A to Semester 2026A, with 34 nights dedicated to Subaru/SCExAO+CHARIS and 8 nights allocated for Keck/NIRC2 at Lp. The goal is to observe approximately 125 stars with magnitudes ranging from V $=
%$3 to 7.2, located at a distance between 20 and 100 pc, and which are aged from $\sim$10 to 800 Myr.

%The program’s goals are to discover, and constrain the atmospheres and orbits of dozens of exoplanets and brown dwarfs but also constrain the atmospheric evolution of substellar objects as a function of mass and time. 
%Furthermore, the survey will provide targets for Nancy Grace Roman(CGI) Technology Demonstrations.

%\textcolor{red}{TC - stopping here 11:59am, 7/16}
\section{Data Sources}

\textbf{Astrometry from the Hipparcos-Gaia Catalogue of Accelerations} -- 
%The {\it Gaia} mission \cite{Gaia_General_2016, Gaia_General_2018,Gaia_Astrometry_2018}, with absolute astrometry for over 1 billion stars, is revolutionizing the field of dynamical mass measurements.  
%indicative of potential companions such as massive planets or low-mass brown dwarfs.
The combination of \textit{Gaia} and the \textit{Hipparcos} mission $\sim$25 years earlier \cite{ESA_1997, vanLeeuwen_2007} provides acceleration measurements for over 115,000 stars.  The \textit{Hipparcos}-\textit{Gaia} Catalog of Accelerations \cite{Brandt_2018} can identify reflex motion from stars with massive planets and brown dwarfs on $\approx$100 year orbits\cite{Brandt+Dupuy+Bowler_2019}.  
%Such long-period systems, often thought to be beyond the reach of dynamical masses, now comprise some of the best candidates for direct imaging follow up.  
Given an angular separation, HGCA data alone give a lower limit to companion mass \cite{Franson2023}, which can be used to remove stars with large accelerations inconsistent with planets or brown dwarfs from our sample.
%identify and remove from further consideration stars accelerated by binary companions, not planets or brown dwarfs.
%The latest version of HGCA incorporates astrometric data from the early third {\it Gaia} data release (EDR3) instead of {\it Gaia} DR2, resulting in a factor of $\approx$4 gain in sensitivity for bright stars \cite{2021A&A...649A...1G,2021A&A...649A...2L} and in turn an increased sensitivity to substellar companions.
%HGCA can reveal accelerations due to a 3\,$M_{\rm Jup}$ on a 30\,au orbit around a star 40\,pc from Earth.
%, as we show in Section \ref{sec:methodology}.  
%This detection criterion corresponds to 3$\sigma$ significance: we can measure the mass of such a planet, on a 160-year orbit, to $\pm$1\,$M_{\rm Jup}$.

\begin{figure}[h]
\centering
\vspace{-0.0in}
\includegraphics[width=1\textwidth]{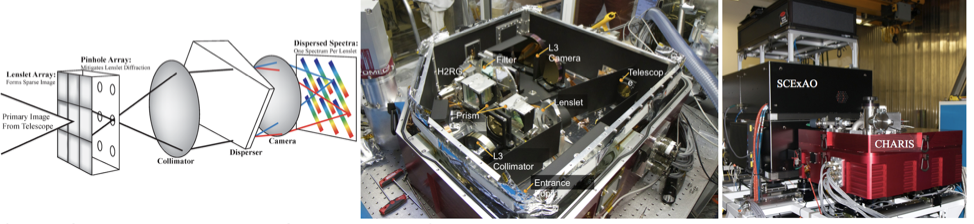}
\caption{(Left) The lenslet array of CHARIS samples the incoming image from SCExAO that is then dispersed into a series of microspectra. CHARIS (Center) is a lenslet-based integral field spectrograph built for the Subaru telescope and located behind SCExAO (Right).}
\label{fig:CHARIS}
\end{figure}

\textbf{Direct Imaging with SCExAO} -- The direct imaging component of our program primarily focuses on the Subaru Coronagraphic Extreme Adaptive Optics project (SCExAO) on the 8.2 m Subaru telescope on Maunakea (Figure \ref{fig:CHARIS}). SCExAO is the world’s leading extreme AO platform in the northern hemisphere, designed to image and characterize jovian planets on solar system scales \cite{Jovanovic2015}. 
Using a 2000-actuator deformable mirror (DM) and a state-of-the-art Pyramid wavefront sensor, SCExAO sharpens starlight partially corrected by Subaru's facility AO system. %188-%actuator facility AO system (AO-188) 
%Augmenting its main wavefront loop, SCExAO uses “empirical orthogonal functions” to predict the future wavefront from past wavefront sensor estimates and accelerometer data \cite{GuyonMales2017}: this ``predictive control'' yields at least a factor of three improvement in SCExAO's raw contrast. 
%Prior to 2024, SCExAO has sharpened starlight partially corrected by Subaru's 188-actuator facility AO system (AO-188) using a 2000-actuator deformable mirror (DM) and a state-of-the-art Pyramid wavefront sensor.  Augmenting its main wavefront loop, SCExAO now uses “empirical orthogonal functions” to predict the future wavefront from past wavefront sensor estimates and accelerometer data \citep{GuyonMales2017}: this ``predictive control'' yields at least a factor of three improvement in SCExAO's raw contrast. 

SCExAO is %utilizes a 2000-actuator deformable mirror driven by a Pyramid wavefront sensor, correcting for up to 1080 modes at up to 3.5 kHz. It is 
coupled with an integral field spectrograph (IFS) named CHARIS  \cite{Peters2012,Groff2016,Groff2017}.
%An IFS allows both images and wavelength information to be taken simultaneously.
CHARIS can yield spatially-resolved spectra at low resolution (R $\sim$ 20) from 1.2 - 2.4 $\mu m$ in a single shot or at a high-resolution of R $\sim$ 70 in the individual J, H, or K-bands. The low-resolution mode is the primary operational mode for planet discovery, as it can clearly reveal the expected sawtooth-like shape of a substellar companion's atmosphere, allowing us to distinguish between a planet/brown dwarf and an unrelated background star in a single data set \cite{Currie2018,Currie2023}.  The high-resolution mode provides follow-up atmospheric characterization for identified candidate companions. To suppress starlight, SCExAO employs multiple coronagraph options (e.g. vector vortex coronagraph, Lyot coronagraph).

%\begin{figure}[h]
%\centering
%\vspace{-0.1in}
%\hspace{-0.4 truein}
%\includegraphics[width=1.0\textwidth]{Figures/contrasts_keck_scexao_v2.pdf}
%\includegraphics[width=0.5\textwidth,trim = 15mm 0mm 0mm 0mm,clip]{images/scexaocharis_contrast.eps}
%\includegraphics[width=0.5\textwidth,trim = 15mm 0mm 0mm 0mm, clip]{images/nirc2_contrast.eps}
%\includegraphics[height=0.385\textwidth]{Figures/scexao_charis_one_hour_contrast_slice_pred.pdf}
%\includegraphics[height=0.385\textwidth]{Figures/nirc2_90min_lp.pdf} \\[0.1em]
%\vspace{-0.1in}
%\caption{Left: AO188+SCExAO/CHARIS can image $\approx$2\,$M_{\rm Jup}$ exoplanets around young, 20 Myr-old stars and 10\,$M_{\rm Jup}$ planets around 300 Myr-old stars: the facility upgrade from AO188 to AO3K improves planet detection capabilities further.    Right: Keck/NIRC2 achieves shallower contrasts in $L^\prime$ than SCExAO/CHARIS in $JHK$, but it can be more sensitive to older, and colder, companions (e.g. those older than 300  Myr) and outperforms competing thermal IR instruments within $0.\!\!^{\prime\prime}$5. }
%\label{fig:scexao_nirc2}
%\end{figure}

\begin{figure}[h]
\centering
\vspace{-0.1in}
\begin{minipage}{0.5\textwidth}
    \centering
    \includegraphics[width=\textwidth,trim = 15mm 0mm 0mm 0mm,clip]{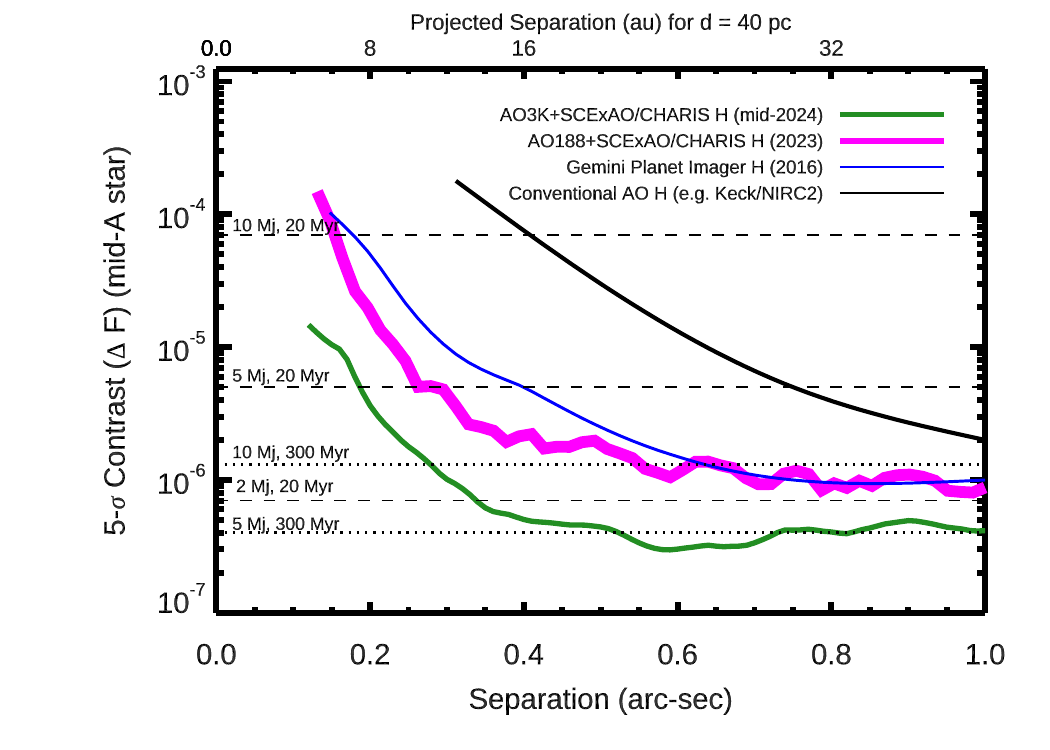}
\end{minipage}%
\begin{minipage}{0.5\textwidth}
    \centering
    \includegraphics[width=\textwidth,trim = 15mm 0mm 0mm 0mm,clip]{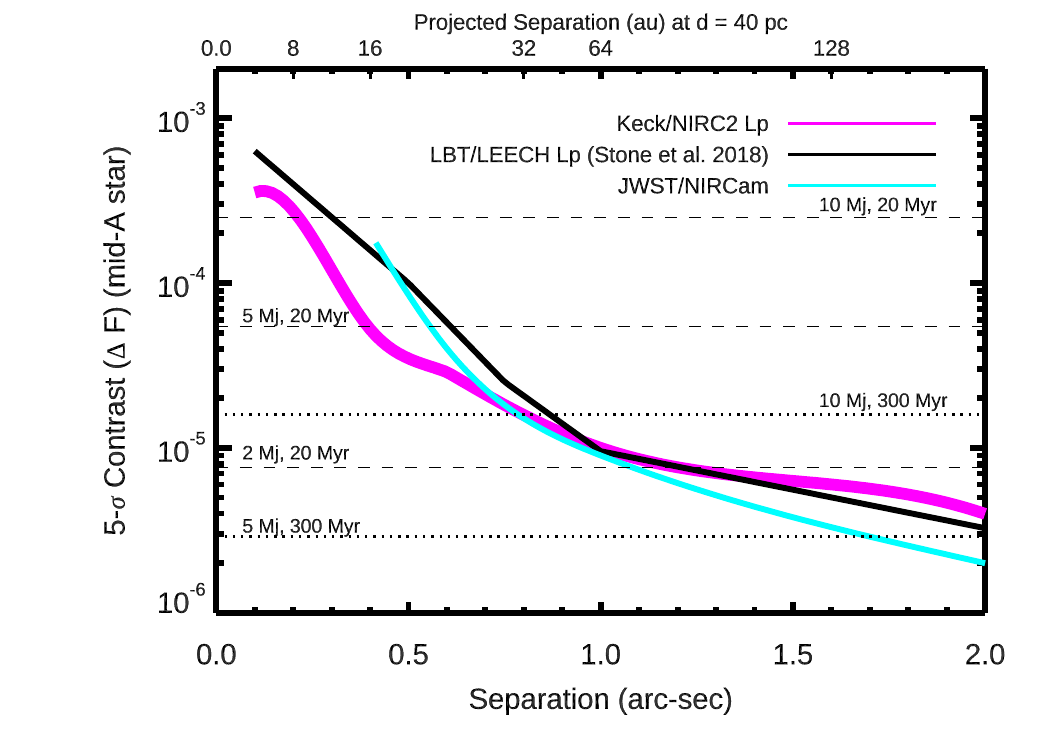}
\end{minipage}
\vspace{-0.1in}
\caption{Left: AO188+SCExAO/CHARIS can image $\approx$2\,$M_{\rm Jup}$ exoplanets around young, 20 Myr-old stars and 10\,$M_{\rm Jup}$ planets around 300 Myr-old stars: the facility upgrade from AO188 to AO3K improves planet detection capabilities further. Right: Keck/NIRC2 achieves shallower contrasts in $L^\prime$ than SCExAO/CHARIS in $JHK$, but it can be more sensitive to older, and colder, companions (e.g. those older than 300 Myr) and outperforms competing thermal IR instruments within $0.\!\!^{\prime\prime}$5. }
\label{fig:scexao_nirc2}
\end{figure}

     %SCExAO/CHARIS are specifically designed for the high contrasts necessary to directly detect extrasolar planets compared to general purpose facilities in the northern hemisphere. 
     For bright stars, SCExAO achieves 2-hour, 5-$\sigma$ contrasts of $\sim10^{-6}$, 3$\times10^{-6}$, and $10^{-5}$ at $0.\!\!^{\prime\prime}$5, $0.\!\!^{\prime\prime}$35, and $0.\!\!^{\prime\prime}$25 (Figure \ref{fig:scexao_nirc2}, left panel).  The long wavelength-baseline offered by CHARIS low-resolution mode enables the use of aggressive, spectral differential imaging (SDI)-based algorithms to achieve deeper contrasts at a closer inner working angles.  With current performance, SCExAO/CHARIS can detect 2--3\,$M_{\rm Jup}$ around nearby A-type stars within $\sim$ 30 au (Figure \ref{fig:scexao_nirc2}, left). 
       %The left panel of Figure \ref{fig:scexao_nirc2} demonstrates SCExAO/CHARIS's sensitivity to young exoplanets down to $\approx$2--3\,$M_{\rm Jup}$ around nearby A-type stars within $\sim$ 30 au.  
       Sensitivities for planets around young, Sun-like stars -- which are intrinsically a factor of 10 or more fainter than A-type stars -- are greater.  E.g. 1 Jupiter-mass planet orbiting at 20 au around a 20 Myr-old Sun-like star at 40 pc (a contrast of $\Delta$F $\sim$ 2$\times$10$^{-6}$) would be detectable.

       SCExAO/CHARIS is now much more powerful thanks to Subaru's upgrade of the facility AO system located immediately upstream. Prior to May 2024, Subaru's facility AO system was AO-188, a 188-actuator DM coupled to a curvature wavefront sensor, which delivered (at best) 20-40\% Strehl at 1.6 $\mu m$, requiring SCExAO's further sharpening to achieve 80-90\% Strehl.  In May 2024, Subaru successfully commissioned their facility AO upgrade: ``AO3K", a 3200 DM with improved wavefront sensing \cite{Guyon2022}.   AO3K by itself yields extreme AO corrections and on-sky contrasts (i.e. a demonstrated 80-90\% Strehl on sky): SCExAO will further sharpen this correction. 
       
       Detailed, wavefront error budget simulations suggest that contrasts achieved with CHARIS behind AO3K’s first-order correction plus SCExAO's 2nd-order correction will be 2--10 times better than those previously achieved with AO-188+SCExAO at $0.\!\!^{\prime\prime}$1--$0.\!\!^{\prime\prime}$5: e.g. 10$^{-6}$ at $0.\!\!^{\prime\prime}$25 and 5$\times$10$^{-7}$ at $0.\!\!^{\prime\prime}$5 (pvt. comm.).   These improved capabilities will allow AO3K+SCExAO/ CHARIS to detect 20 Myr-old 2-Jupiter-mass planets on Saturn-like orbits around young, A-type stars and sub-Jupiter-mass planets around 20 Myr-old Sun-like stars.  %However, \textbf{for this proposal, we emphasize that our expected program yield goals are fully met by prior performance with AO188+SCExAO/CHARIS: the AO3K is ``value added" beyond our requirements} (see Section \ref{sec:yield}).

\textbf{Keck/NIRC2 Thermal IR Imaging} -- We complement our SCExAO/CHARIS imaging program with $L_{\rm p}$ data from the NIRC2 camera on the on Keck II Telescope.   NIRC2 enables imaging at longer wavelengths than CHARIS (3--5 $\mu m$), where older and/or colder objects are brighter \cite{Skemer2014} .  
%NIRC2 in $L^\prime$ %with the pyramid wavefront sensor 
%can also detect young exoplanets around nearby stars (Figure \ref{fig:scexao_nirc2}, right panel).  
For ages of about 300 Myr (greater than 300 Myr), Keck/NIRC2 is roughly as sensitive (more sensitive) to planets as the pre-upgrade SCExAO/CHARIS (Figure \ref{fig:scexao_nirc2}, right panel).  

We will use the upgraded SCExAO/CHARIS for challenging targets at young ages and small angular separations, where it is more sensitive than NIRC2.  NIRC2 will be competitive with SCExAO/CHARIS around older stars.  We will use the performance of both instruments shown in Figure \ref{fig:scexao_nirc2}, combined with knowledge of the system age (see next section) to choose whether to observe targets with SCExAO/CHARIS or NIRC2.  We will follow up our discoveries with both instruments to obtain full spectral and wavelength coverage.

\begin{figure*}
 \centering
 \includegraphics[width=\textwidth]{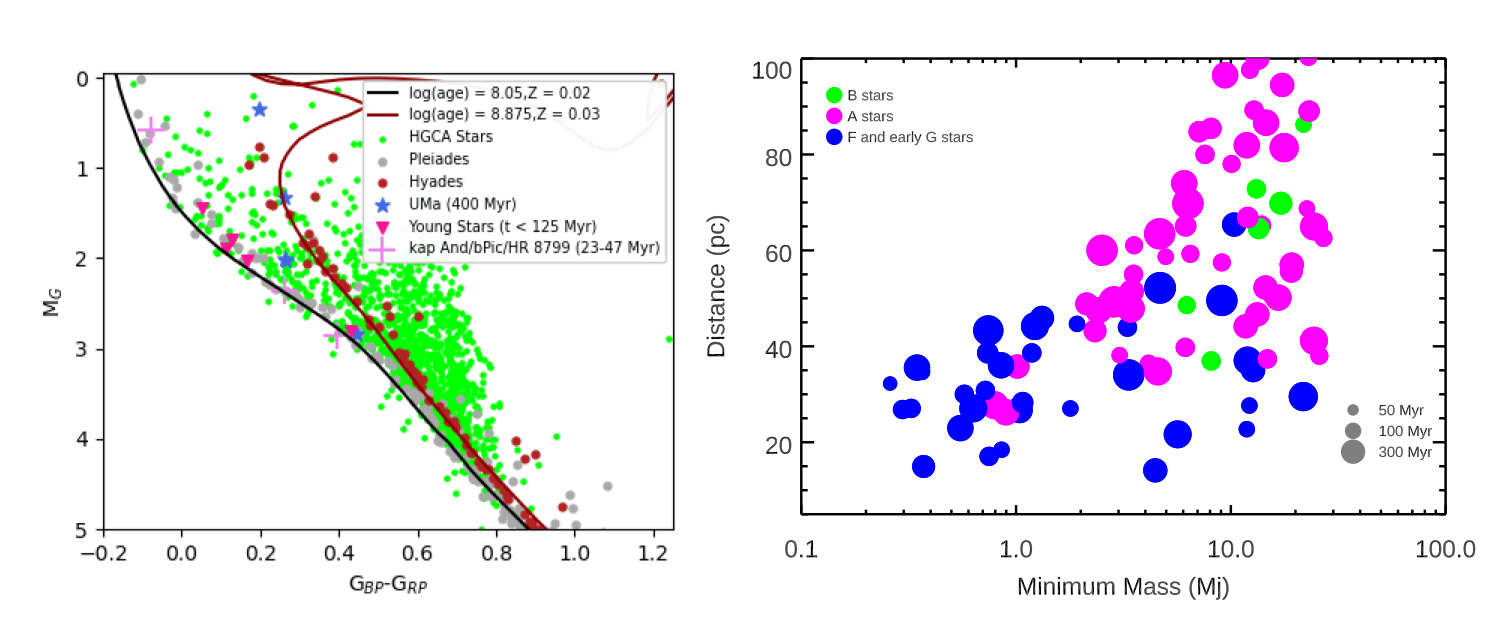}
% \caption{Color-magnitude diagram using Gaia DR3
%photometry and displaying PARSEC isochrones for the Pleiades (115 Myr; dark gray line) and Hyades (750 Myr; red line). Reproduced from Tobin et al. \cite{Tobin2024}.}
\caption{ HR diagram comparing the positions for all BAF-type accelerating stars to the Pleiades, Hyades, very young objects (23--47 Myr), other young stars with inteferometrically derived ages of less than 125 Myr, and Ursa Majoris members ($\sim$400 Myr). Rotation effects may make some early-type stars comparable to the Pleiades in age or even younger lie above the Pleiades locus. We use other age diagnostics (e.g. CHARA interferometric data of the star; moving group membership) to augment our age estimates. (right) Distance vs. minimum mass at 0\farcs{}2 for a subset of our sample: symbol sizes are proportional to estimated age.   The minimum mass scales as $\rho^{2}$: e.g. 4 $M_{\rm J}$ at 0\farcs{}2 = 9 $M_{\rm J}$ at 0\farcs{}3 or 20 $M_{\rm J}$ at 0\farcs{}45.}
 \label{fig:TS}
\end{figure*} 

\section{Target Selection}
\label{sec:TS}

We focus our selection on young stars exhibiting evidence of astrometric acceleration plausibly due to an unseen but potentially imageable planet or brown dwarf. To define the significance threshold in acceleration for including a star in our sample, we use results from our recent pilot survey\cite{Currie2021}.  There, we discovered two companions around stars accelerating at $\sim$2.2$\sigma$ \cite{Currie2023}(T. Currie in prep.) and found no clear evidence that our detection rate was higher (or false positive rate lower) if we only considered the strongest accelerators (i.e. $>$3--5$\sigma$).  Therefore, our sample includes stars that are accelerating at a $\ge$2.2-$\sigma$ significance.  For comparison, only 15\% of GPIES targets and 16\% of LEECH targets show a $>$2.2$\sigma$ acceleration from HGCA.  

Furthermore, we focus on stars whose accelerations can plausibly be due to a planet or brown dwarf at angular separations accessible by our direct imaging observations (0\farcs{}1--1\arcsec{}).  From Brandt et al. \cite{Brandt+Dupuy+Bowler_2019}, acceleration is related to the companion mass: $a_{\alpha\delta} = \frac{GM_{b}}{r^{2}}cos\phi$.   Thus, assuming a perfectly face-on system ($\phi$ = 0) where the companion's orbit is significantly longer than the orbital time, we can estimate a lower limit to the companion mass if we know the angular separation. We select accelerating systems with a minimum companion mass of $M_{\rm min}$ $<$ 30 $M_{\rm Jup}$ at $\rho$ = $0.\!\!^{\prime\prime}$2.  As confirmed by the pilot survey, systems with higher values of $M_{\rm min}$(0\farcs{}2) harbor stellar companions or planets at much smaller separations inaccessible to SCExAO and Keck.

%Our selection criteria for identifying accelerators is informed by the results of a pilot survey (see Section \ref{sec:yield}). The pilot survey focused on stars accelerating at a 2-$\sigma$ to 50-$\sigma$ level. We discovered two companions around stars accelerating at $\sim$2.2$\sigma$ and found no clear evidence that our detection rate was higher (false positive rate lower) if we only considered the strongest accelerators (i.e. $>$3--5$\sigma$).  Therefore, we select stars that are accelerating at $\ge$2.2-$\sigma$: similar to our pilot survey.  The HGCA has demonstrated an accurate calibration of uncertainties; $\approx$30\% of stars show evidence of acceleration. %We will prioritize targets with a $>$3-$\sigma$ acceleration.

%The acceleration is related to the companion mass: $a_{\alpha\delta} = \frac{GM_{b}}{r^{2}}cos\phi$. 
%We conservatively assume each companion's orbit is viewed face-on, although the expected $<cos\phi>$ = 0.5. 
%We will reject systems with a minimum mass of $M_{\rm b}$ $>$ 30 $M_{\rm Jup}$ at $0.\!\!''2$. 
%masses harbor stellar companions or planets at much smaller separations inaccessible to SCExAO and Keck (pvt. comm.).

Direct imaging favors planet detection around young stars because exoplanets (and brown dwarfs) contract, cool, and therefore fade with time\cite{CurrieBiller2023}.
%Exoplanets and brown dwarfs cool and fade with time: our detection capabilities vs.~age thus favor younger stars.
Thus, we pinpoint the subset of accelerating stars that are young enough to have direct imaged companions according to planet evolution models.  We select for youth primarily based on moving group membership, \textit{Gaia} HR diagram positions, or interferometry (Figure \ref{fig:TS}).  We focus on B, A, and early F stars because a) their ages can be better inferred from HR diagrams (e.g. \cite{Tobin2024}), b) they have a higher frequency of imaged planets \cite{Nielsen2019}, and c) their optical brightnesses enable higher Strehl ratios and make the targets more relevant for future NASA missions, in particular Roman-CGI. 
%Early analysis shows that age estimates -- considered as an ensemble -- are precise enough for our science goals (e.g. to inform planet luminosity evolution vs mass; Sect 4.1). 
%We use this information to focus on stars whose unseen companions could be imageable. 
%\textit{Gaia} photometry help identify young BAF stars: we select targets whose HR diagram positions are closer to the Pleiades (115 Myr) than the Hyades (700 Myr), inspecting targets manually for ancillary age information (e.g. moving groups, CaHK [18]). 
The frequency of Jovian planets drops significantly beyond 30 au \cite{Fulton2021}: 
%At a given angular separation, nearby stars probe smaller physical projected separations.  
we set a distance cutoff of 100 pc, so that $0.\!\!''3$ probes $r_{\rm proj}$ $<$ 30 au.  Finally, we use archival data including the Washington Double Stars Catalog and the Keck Observatory Archive to remove targets with known stellar companions.

 From this process, we obtain a sample consisting of 175 stars. About 40\% of the targets have V $\le$ 5; most are brighter than V = 6.  Most lie within 60 pc with a median age of 150 Myr: minimum companion masses at $0.\!\!^{\prime\prime}$2 are 0.3--30 $M_{\rm Jup}$  (median = 7 $M_{\rm Jup}$) (Figure \ref{fig:TS}). From this list, we will observe 125 stars, chosen based on observing schedules for a baseline of 2 hours of integration time per target.   The majority of these targets have never been observed with extreme AO systems before: most had no public ground-based AO or \textit{Hubble Space Telescope} observations either.
 %For our survey, we plan to observe 125 of these targets, which 25 have V $<$ 5-7.2 with a median estimated age of 150 Myr. From this target list, 90 chosen targets will be observed contingent upon telescope schedules and other logistical factors.

\begin{figure*}
 \centering
 \includegraphics[width=0.9\textwidth]{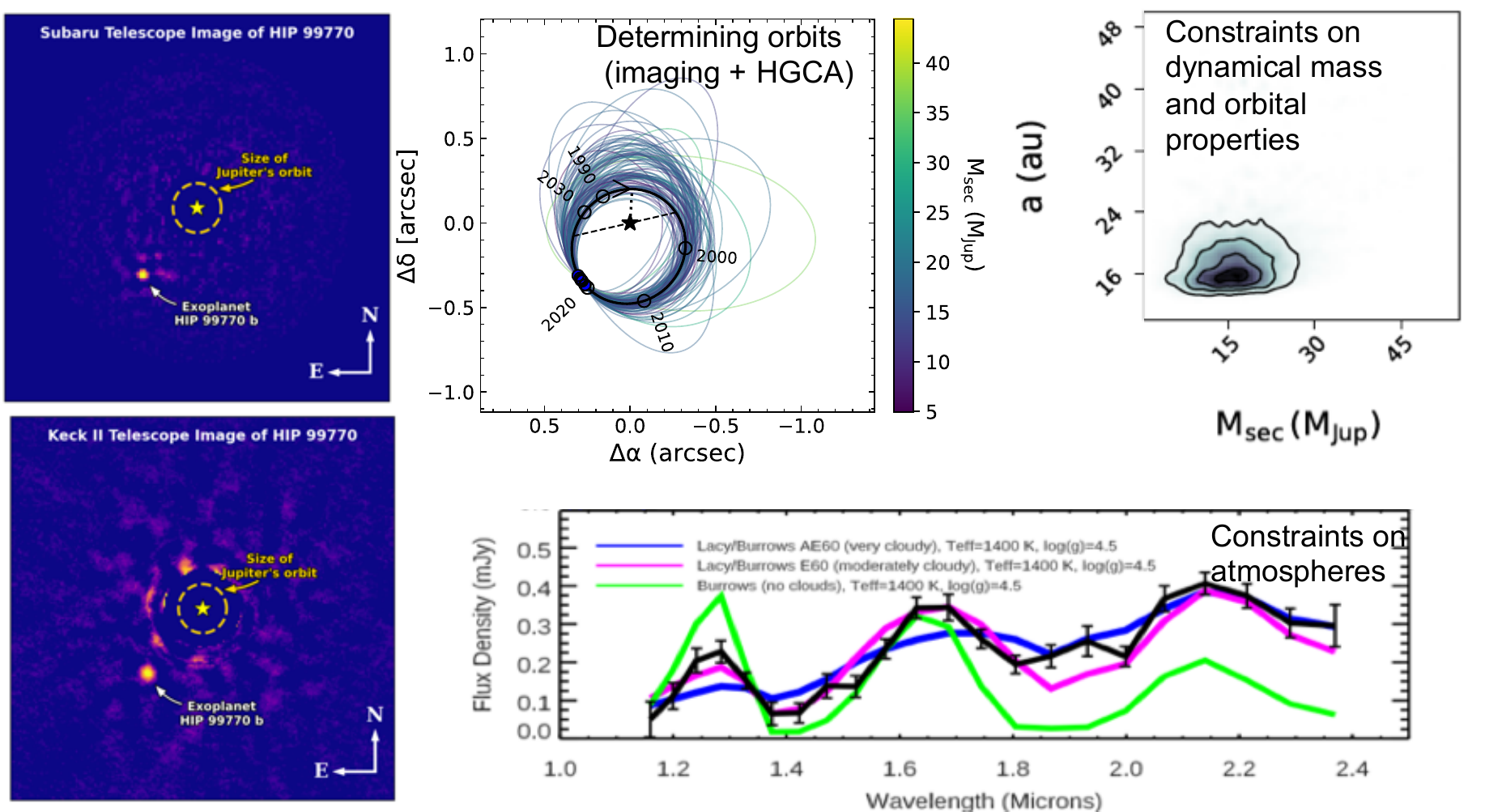}
 \caption{ Discovery and characterization of HIP 99770 b from our pilot survey.   (left) Around the accelerating star HIP 99770, we detect a companion with SCExAO/CHARIS and Keck/NIRC2.  We model its astrometry to constrain its mass and orbit (top-middle, top-right).   Comparing its SCExAO/CHARIS near-IR spectra and Keck/NIRC2 thermal IR photometry yields constraints on the planet's temperature, surface gravity, and clouds (bottom-right). }
 \label{fig:poc}
\end{figure*}% 

\section{Proofs of Concept for this Survey, Predicted Yield of Discoveries, and Preliminary Results}
\label{sec:first_results}

%Our survey started in February 2024, and we have currently completed approximately 8\% of the planned observations. By the end of 2024, we anticipate that around one-third of the survey will be finished. Already, the survey has yielded significant findings, including the confirmation of a probable new planet, the verification of two candidate brown dwarfs, and the identification of two additional candidate substellar companions.

%\noindent Based on the results from our pilot study and preliminary data from the Intensive Survey, we predict that we will discover, determine the masses, and constrain the orbits of five planets and twelve brown dwarfs. We expect our discovery yield to be five times higher than that of larger, unbiased surveys, such as the Gemini Planet Imager Exoplanet Survey (GPIES).

HIP 99770 b, the first planet discovered through both direct imaging and astrometry\cite{Currie2023}, serves as a proof-of-concept for our survey (Figure \ref{fig:poc}).  We identified HIP 99770 as a young, accelerating star from HGCA and HR diagram analysis.  Follow-up SCExAO/CHARIS data at 1.1--2.4 $\mu m$ obtained over 18 months imaged a companion -- HIP 99770 b -- at a projected separation of $\rho$ 0\farcs{}43--0\farcs{}44.   Follow-up Keck/NIRC2 data re-detected HIP 99770 b at 3.8 $\mu m$. Using orvara \cite{Brandt2021} to model HIP 99770’s data allowed for precise orbital constraints, with the semimajor axis estimated at about 16.9 au with 15\% precision. Its mass is estimated at 16.1$^{+5.4}_{-5.0}$$M_{J}$ using conservative priors, or 13.9$^{+6.1}_{-5.1}$ $M_{J}$ with a log-normal prior. Atmospheric modeling shows that HIP 99770 b is a substellar object near the L/T transition with clouds intermediate in thickness between the youngest planets like HR 8799 bcde and field brown dwarfs.  Likewise, the discovery of AF Lep b through direct imaging and astrometry demonstrates the efficacy of a planet search whose target list is comprised of accelerating stars \cite{deRosa2023} and the ability of such surveys to identify companions down to $\approx$2--4 $M_{\rm J}$.
%with an estimated quite cloudy planet with a temperature of 1400 K, indicating an intermediate level of cloudiness compared to the youngest imaged planets.

Based on the results of our pilot survey conducted over the past 5 years, we predict a planet and brown dwarf discovery yield higher than that larger blind surveys like GPIES.  The pilot survey targeted about 50 stars with typical integration time of 30-60 minutes, or 2-3 times shorter than in this program.    Other factors further reduced the pilot's effectiveness: e.g. the Kilauea eruption compromised SCExAO/CHARIS's performance at first (soot on the IR secondary).  Beyond that, the pilot survey's target selection strongly mirrored those in this proposal in age and distance. 

In addition to the published planet around HIP 99770, the planet or brown dwarf around HIP 39017 \cite{Tobin2024}, and published brown dwarfs like HIP 21152 B \cite{Kuzuhara2022}, the survey identified other substellar companions, which will be published at a later date.  In total, the pilot achieved a substellar companion discovery rate of about 16\%.   Given our survey's sample size (twice that of the pilot), increased depth per target (2-3$\times$ increase in integration time), and upgraded hardware yielding deeper contrasts per observation (i.e. the AO3K upgrade), we expect our survey will result in a total yield of 5 new exoplanets and 12 brown dwarfs with well constrained atmospheres, orbits, and dynamical masses.  The total yield of new detections will be more than 5 times higher than blind surveys such as GPIES despite having a smaller sample size.
%{\bf we should discover at least 5 new exoplanets that are imaged, weighed, have their atmospheres characterized, and have their orbits tracked.  We should also discover $\ge$12 new brown dwarfs that can be characterized in a similar way.  Our total yield will therefore be about 5 times higher than GPIES.}  

%\textbf{Given this proposal's larger sample, greater depth per target, and upcoming improved performance from to the AO3K upgrade, the pilot survey results provides the basis for a conservative estimate of our predicted yield.}

%A pilot SCExAO survey has observed 50 targets so far: only 25 of these at the 2-hour depth we plan going forth.  
%{\bf This pilot program discovered 2 planets (HIP 99770 b \cite{Currie2023} and one unpublished), a substellar companion that likely a planet (HIP 39017 b; \cite{Tobin2024}; Figure \ref{fig:newdisc}, left panel), and 5 brown dwarfs (3 published)}: a planet discovery rate of $\sim$4--6\% and planet+brown dwarf detection rate of 16\%.  {\bf This is over $\sim$10 times higher than the discovery rates for planets and substellar companions from GPI} \cite{Nielsen2019}, which yielded only one new planet discovery and two new brown dwarf discoveries out of 300 targets.    Given this survey's sample size (twice that of the pilot), {\bf we should discover at least 5 new exoplanets that are imaged, weighed, have their atmospheres characterized, and have their orbits tracked.  We should also discover $\ge$12 new brown dwarfs that can be characterized in a similar way.  Our total yield will therefore be about 5 times higher than GPIES.}  

\begin{figure*}
 \centering
 \includegraphics[width=\textwidth]{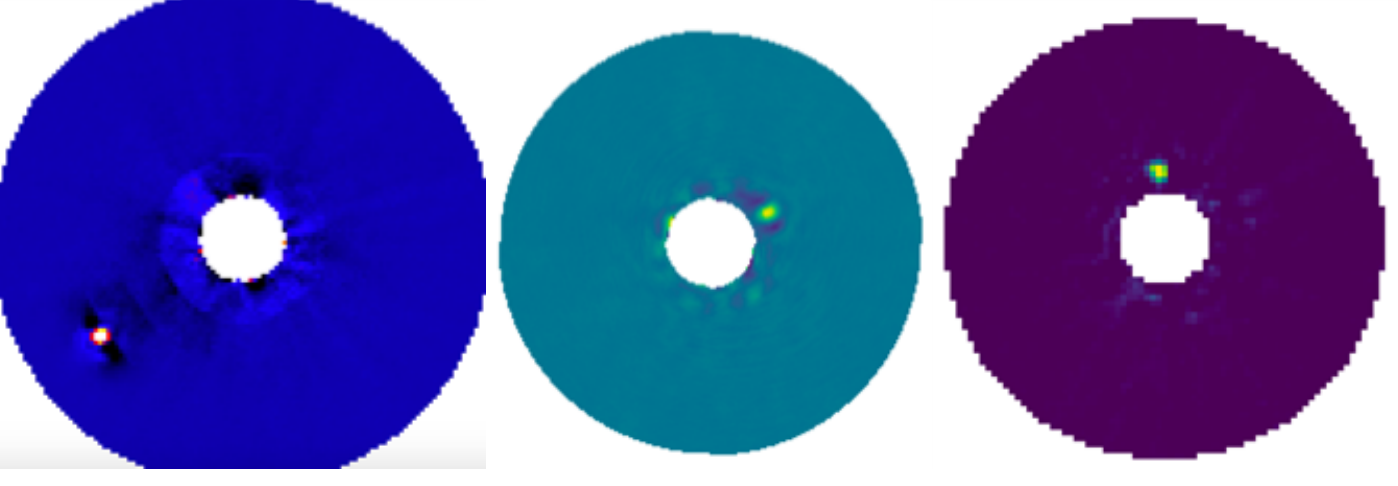}
 \caption{Selected new detections of low-mass companions from our survey data obtained during the first semester.   SCExAO/CHARIS data during the second half of 2024 will be obtained with improved sensitivity thanks to the AO3K upgrade, making our survey more sensitive to detecting faint planets at small angular separations.}
 \label{fig:first_results}
\end{figure*}

Our survey started in February 2024, and as of 16 July 2024 we have currently completed approximately 13\% of the planned observations.   Starting in August, we expect to carry out SCExAO/CHARIS observations in combination with the AO3K upgrade.  By the end of 2024, we anticipate that nearly one-third of the survey will be finished.  Already, we have identified a confirmed (likely) planet, at least one new confirmed brown dwarf, and multiple other candidate low-mass companions (Figure \ref{fig:first_results}).

%\begin{figure*}
 % \centering
 % \includegraphics[width=0.6\textwidth]{implementation_fim.pdf}
 % \caption{Schematic of the design of the HiRISE/FIM. Reproduced with permission from El Morsy et al. \cite{mel2022}.}
 % \label{fig:fim}
%\end{figure*} 

\section{Follow-up characterization of known planets and brown dwarfs around accelerating stars}
\label{sec:follow-up}
As demonstrated by the many recent studies of AF Lep b, a planet or brown dwarf discovered from an accelerating star survey is well-suited for extensive follow-up atmospheric and dynamical characterization (e.g. \cite{Gratton2024,Franson2024}).   We can perform follow-up observations of companions discovered from our program with CHARIS at higher resolution across individual J, H, and K bands. These observations allow us to more precisely investigate surface gravity and chemical composition. 
%Follow-up characterization studies of the planets and brown dwarfs identified during the pilot survey are currently underway, providing deeper insights into their physical and atmospheric properties.

\begin{figure*}
 \centering
 \includegraphics[width=0.8\textwidth]{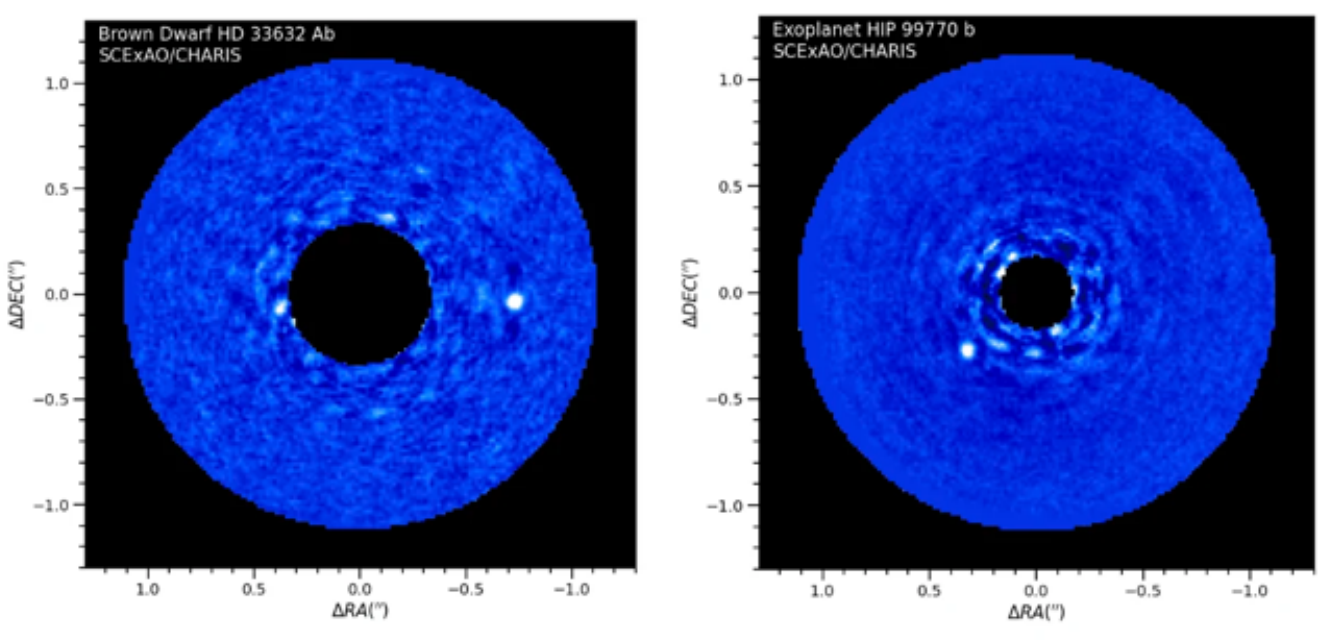}\\
  \includegraphics[width=0.7\textwidth]{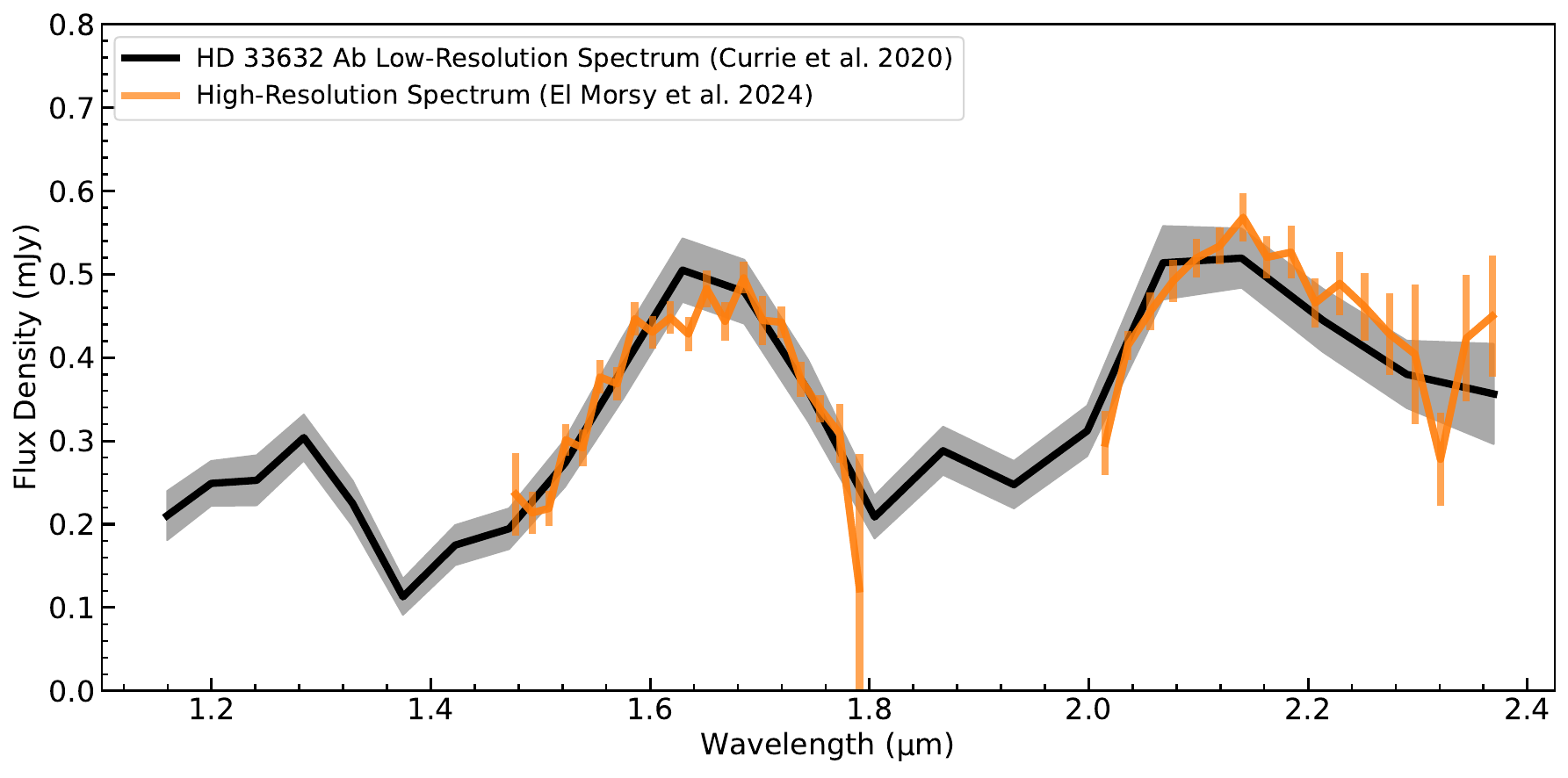}
 \caption{Follow-up CHARIS data for substellar companions identified from our pilot survey.   (top panel) CHARIS K Band images of HD 33632 Ab (left; \cite{ElMorsy2024}) and HIP 99770 b (Bovie et al. 2024, in preparation).  (bottom panel) The $H$ and $K$ band spectra for HD 33632 Ab compared to lower-resolution spectra\cite{Currie2020}.}
 \label{fig:FU}
\end{figure*} 
As an example, Figure \ref{fig:FU} shows detections of the brown dwarf HD 33632 Ab and planet HIP 99770 b obtained with SCExAO/CHARIS in the $H$ and $K$ bands (R $\sim$ 70).   These companions were identified from prior pilot-survey observations with SCExAO/CHARIS\cite{Currie2020,Currie2023}.  While new spectra for these companions are consistent with lower-resolution spectra from the discovery papers, these new data provide an improved constraint on gravity and chemistry-sensitive features (El Morsy et al. 2024 submitted; Bovie et al. 2024, in preparation).   New measurements of the companions' positions from imaging data and complementary data (e.g. SOPHIE spectrograph radial-velocity data for HD 33632 Ab) help to more precisely determine the objects' orbital properties and dynamical masses.
%These new data are provided alongside astrometry from HCGA and additional radial-velocity observations from the SOPHIE spectrograph.

%Fig. \ref{fig:spectra} displays the spectra of HD 33632 Ab in the $H$ and $K$ bands obtained with CHARIS in October 2021 (plotted in magenta) are compared to lower-resolution spectra from CHARIS taken in 2020 by Currie2020a (plotted in black). The vertical bars represent the errors.

%\begin{figure*}
% \centering
% \includegraphics[width=0.7\textwidth]{images/hd33632spectrumcomp.pdf}
% \caption{Follow-up SCExAO/CHARIS spectrum of HD 33632 Ab obtained in the $H$ and $K$ bands at a resolution of $R$ $\sim$ 70 (El Morsy et al. 2024, submitted) compared to the lower-resolution spectrum from the discovery paper.}
% \label{fig:spectra}
%\end{figure*} 

\section{Significance for the {Roman Space Telescope} {Coronagraphic Instrument}}
This program has significant implications for and was in-part allocated time to support the \textit{Roman Space Telescope} Coronagraphic Instrument by retiring a significant risk for Roman-CGI's technology demonstration program\cite{Kasdin2020}. 

\subsection{Roman-CGI: Suitable Tech Demo Targets are Needed but Missing}
The Roman-CGI tech demo is a critical stepping stone towards a future NASA flagship mission able to achieve the ultimate goal of directly imaging and characterizing an Earth-like planet in reflected light.
%The CGI tech demo is a critical stepping stone towards a future NASA flagship mission able to achieve this ultimate goal.
The NASA-approved criteria for the CGI tech demo's success\footnote{ \scriptsize{\url{https://roman.gsfc.nasa.gov/science/rsig/2021/Roman_Requirements_20201105.pdf}}} consists of five Objectives (2.2.1--2.2.5) and one Threshold Technical Requirment (TTR5), which is a Level 1 Requirement.   
%Some Objectives focus purely on technical details -- e.g. demonstrating use of advanced coronagraphs/coronagraph software, modifying wavefront control algorithms in flight (2.2.2, 2.2.3). 
Objective 2.2.1 and 2.2.5 require detecting companions around two stars \textit{``at a contrast level and separation that requires a functional coronagraph and wavefront control capability"}  and characterizing ``\textit{photometry, spectroscopy, and astrometry"} of at least one of them.
%characterized
%; Objective 2.2.5 further requires photometric, astrometric, and spectroscopic measurements of at least one point source. 
TTR5 explicitly states that CGI \textit{must} detect at $>$5-$\sigma$ a point source at a contrast\footnote{
%The language of TTR5 could be misinterpreted to imply that a 5$\sigma$ detection of, say,  a 5$\times$10$^{-7}$ contrast source would fulfill requirements.  
The intent of TTR5 is ``detect a point source that is 10$^{-7}$ times the star's brightness \textit{or fainter}".  E.g. a 5$\times$10$^{-8}$ contrast source would fufill this requirement.} of at least 10$^{-7}$ at $\lambda_{\rm c}$ $\le$ 600 nm ($>$10\% bandpass) located 6--9 $\lambda$/D ($\sim$ 0.3"--0.45") from a very bright (V$_{\rm AB}$ $\le$ 5) star.  

However, {the peer-reviewed literature currently lacks any imaged exoplanets demonstrably satisfying all three of these  requirements}\footnote{The CGI Design Reference Mission does contemplate that TTR5 could be fulfilled ``by analysis" -- a companionless star and simulated planets.  But arguably Objective 2.2.1 and 2.2.5 will remain unsatisfied in this case, and  star \textit{with} a companion would be a significant advantage for CGI.}. 
$\beta$ Pic bc are undetectable at 575 nm due to the system's debris disk;
%is unsuitable for planet detection due to its very bright debris disk;
%\footnote{$\beta$ Pic's extremely bright debris disk will likely drown the signals of $\beta$ Pic bc with CGI at 600nm [2].}, 
published models suggest that 51 Eri b is too faint and HR 8799 e is potentially too faint\cite{Lacy2020}.  All other known imaged planets orbit stars that are fainter than V=5 and/or are located exterior to 6--9 $\lambda$/D, most are outside CGI's control radius.  CGI wavefront control should work at V=6 for the tech demo but likely degrades for fainter stars\cite{Shi2017}.   Even worse, new iterations of atmosphere models first presented in \cite{Lacy2020} predict \textit{fainter} self-luminous planet brightnesses at 575 nm: e.g. likely putting HR 8799 e out of reach (T. Currie, 2024 unpublished).
%Even considering far fainter stars, known imaged systems provide CGI with very few targets to choose from. 

A focus only on mature, radial-velocity (RV) detected Jupiter-mass planets only exacerbates these problems.
%does not solve these problems.
With few exceptions, well-characterized mature planets around V $<$ 5 stars detected by RV lie interior to 3--6 $\lambda$/D for CGI at 575 nm or require steep contrasts $\lesssim$10$^{-9}$ for detection (e.g. $\upsilon$ And): only feasible if CGI's current best laboratory contrasts are realized in flight and 100x more challenging than the 10$^{-7}$ contrast requirement \footnote{RV systems must also be targeted only when the planets' orbits put them at $\rho$ = 0\farcs{}3--0\farcs{}45.  A search of potential targets with EXOSIMS reveals that, with few exceptions, the orbits are not yet well characterized enough to predict when this occurs.  Other RV-detected companions that may be within the CGI dark hole will be too faint to be detectable (T. Currie, 2024 unpublished).  }.  The dearth of demonstrated CGI-accessible planets also affects any CGI science program following a successful CGI tech demo since a science program is nominally contingent upon a successful tech demo result.

\begin{figure*}
 \centering
 \includegraphics[width=0.7\textwidth]{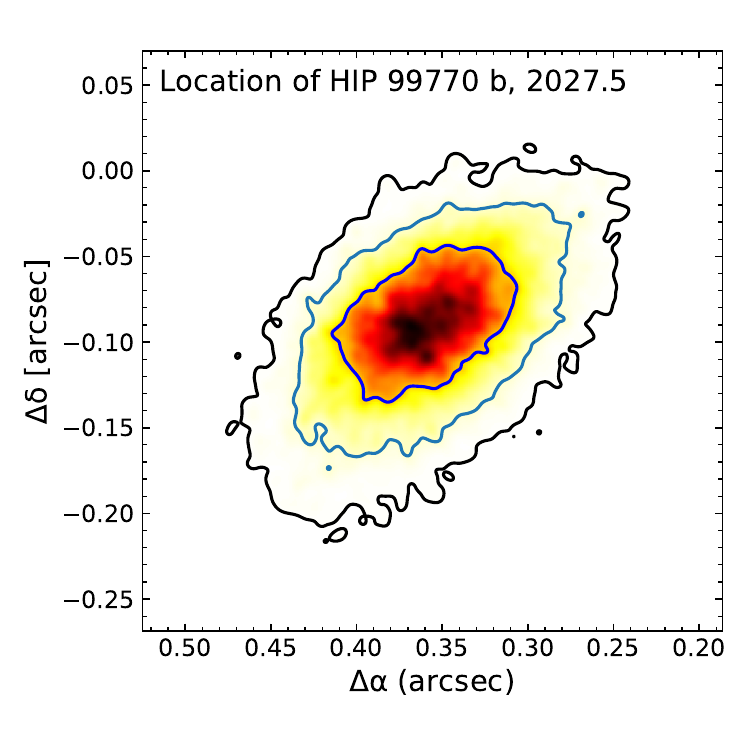}
 \caption{Predicted location of HIP 99770 in July 2027 based on the dynamical modeling from Currie et al \cite{Currie2023}.   HIP 99770 b will likely lie within the Roman CGI dark hole.   The primary is brighter than V = 5.  If the companion is sufficiently bright at 575 nm, it would be suitable as a CGI tech demo target.}
 \label{fig:cgipred}
\end{figure*} 

\subsection{Identifying Roman-CGI Tech Demo Targets From This Program}

We will use the best-fit atmospheric parameters derived from CHARIS and NIRC2 -- e.g. $T_{\rm eff}$, log(g), clouds -- to coarsely predict optical fluxes and planet-to-star contrasts at 600nm and other CGI passbands, determining where companion contrasts fall
%, assuming a range of values for parameters not well probed by CHARIS or NIRC2 that are well diagnosed in the optical (i.e. metallicity).  
%Our goal is to \textit{estimate} optical contrasts, so we only need a coarse prediction of a companion's optical brightness to determine where it falls 
within the 10$^{-7}$ to 10$^{-9}$ range bracketing CGI's likely achievable performance.
%\footnote{As described in Sect. 1.5, planets and brown dwarfs this survey is sensitive to will likely have contrasts falling in this range}.
We will generate a catalogue of planets and low-mass brown dwarfs that 1) fulfill CGI tech demo requirements as written, 2) would fulfill requirements if V can be 6--6.5 mag, 3) would fulfill requirements of angular separation criteria are relaxed (e.g. $\rho$ = 0\farcs{}2--0\farcs{}45).   

To assess whether our companions will lie within the Roman-CGI dark hole during the likely tech demo execution, we will predicted companion locations in 2027--2028 based on their best-fitting orbital parameters from \texttt{orvara}.  Figure \ref{fig:cgipred} shows an example analysis, demonstrating that HIP 99770 b will likely lie within the Roman-CGI dark hole region early in the Roman mission (likely during the tech demo phase).   Overall, we find three published companions from our pilot program -- HIP 99770 b, HIP 21152 B, and HIP 39017 b -- that will be within the CGI dark hole region in 2027-2028 and multiple unpublished companions from our survey that may lie within this region.  

Beyond populating the Roman CGI tech demo target list, this program may enhance the science return of CGI in general. Companions that will be too faint at 575 nm for the CGI tech demo may nevertheless be bright enough at redder wavelengths for follow-on CGI science observations\cite{ElMorsy2024}.  A large population of exoplanets with high-quality spectra, well constrained orbits, and dynamical mass measurements will provide crucial insights into how the atmospheres of jovian planets with different masses evolve with time.

\acknowledgments % equivalent to \section*{ACKNOWLEDGMENTS}       
This work is supported by NASA-Keck Strategic Mission Support program 80NSSC24K0943.
 
% References
\clearpage
\bibliography{report} % bibliography data in report.bib
\bibliographystyle{spiebib}

\end{document}